\documentclass[fleqn,usenatbib]{mnras}

\usepackage{newtxtext,newtxmath}

\usepackage[T1]{fontenc}
\usepackage{ae,aecompl}

\usepackage{graphicx}	
\usepackage{amsmath}	
\usepackage{amssymb}	
\graphicspath{{./}{figures/}}
\usepackage{color,subfigure,lineno} 
\usepackage{amsmath}
\usepackage{array}
\usepackage{soul}
\let\oldAA\AA
\renewcommand{\AA}{\text{\normalfont\oldAA}}
\usepackage{rotating}
\usepackage{color}

\title[Extreme Variability Quasars]{Characterization of Optical Light Curves of Extreme Variability Quasars Over a $\sim 16$-yr Baseline}
\author[Luo et al.]{Yuanze Luo,$^{1}$
Yue Shen,$^{1,2}$\thanks{E-mail: shenyue@illinois.edu}
Qian Yang$^{1}$
\\
$^{1}$Department of Astronomy, University of Illinois at Urbana-Champaign, Urbana, IL 61801, USA \\
$^{2}$National Center for Supercomputing Applications, University of Illinois at Urbana-Champaign, Urbana, IL 61801, USA \\
}

\date{Accepted XXX. Received YYY; in original form ZZZ}

\pubyear{2019}

\begin{document}
\label{firstpage}
\pagerange{\pageref{firstpage}--\pageref{lastpage}}
\maketitle

\begin{abstract}
We study the optical light curves -- primarily probing the variable emission from the accretion disk -- of $\sim 900$ extreme variability quasars (EVQs, with maximum flux variations more than 1 mag) over an observed-frame baseline of $\sim 16$ years using public data from the SDSS Stripe 82, PanSTARRS-1 and the Dark Energy Survey. We classify the multi-year long-term light curves of EVQs into three categories roughly in the order of decreasing smoothness: monotonic decreasing or increasing ({3.7}\%), single broad peak and dip ({56.8}\%), and more complex patterns ({39.5}\%). The rareness of monotonic cases suggests that the major mechanisms driving the extreme optical variability do not operate over timescales much longer than a few years. Simulated light curves with a damped random walk model generally under-predict the first two categories with smoother long-term trends. Despite the different long-term behaviors of these EVQs, there is little dependence of the long-term trend on the physical properties of quasars, such as their luminosity, BH mass, and Eddington ratio. The large dynamic range of optical flux variability over multi-year timescales of these EVQs allows us to explore the ensemble correlation between the short-term ($\lesssim 6$ months) variability and the seasonal-average flux across the decade-long baseline (the rms-mean flux relation). We find that unlike the results for X-ray variability studies, the linear short-term flux variations do not scale with the seasonal-average flux, indicating different mechanisms that drive the short-term flickering and long-term extreme variability of accretion disk emission. Finally, we present a sample of {16} EVQs, where the approximately bell-shaped large amplitude variation in the light curve can be reasonably well fit by a simple microlensing model.    
\end{abstract}

\begin{keywords}
galaxies: active  -- quasars: general -- black hole physics
\end{keywords}


\section{Introduction} \label{sec:introduction}

Variability is a hallmark signature of accreting supermassive black holes (SMBHs) across a broad range of wavelengths, timescales, and accretion states \citep[e.g.,][]{Mushotzky_etal_1993,Ulrich_etal_1997,Peterson_2001}. In the rest-frame UV-optical, the emission and its variability from quasars -- luminous, efficiently accreting SMBHs -- primarily come from the accretion disk, and variability probes the structure of the accretion disk and constrains the physical mechanisms of accretion processes, where the emitting region is too compact to be spatially resolvable.   

The study of quasar variability in the optical band has been greatly advanced in the past decade. There have been several major optical imaging surveys that provide high-quality and long-baseline light curves for large statistical samples of quasars across wide redshift and luminosity ranges \citep[e.g,][]{York_etal_2000,Chambers2016,Flaugher_2005}. These light curve data have enabled the most comprehensive measurements of the optical variability of quasars as functions of timescales, luminosity, color, and BH mass \citep[e.g.][]{VandenBerk2004,MacLeod_etal_2010,MacLeod2012,Schmidt_etal_2010,Simm_etal_2016}. 

A notable result from these optical variability studies is the discovery of hypervariable quasars, which can vary by more than one magnitude over multi-year timescales, compared to the typical variability amplitude of $\sim 0.2$ mag on these timescales \citep[e.g.,][]{MacLeod_etal_2010}. While the existence of this rare population has long been known based on the monitoring of low-redshift Seyferts \citep[e.g.,][]{Penston_Perez_1984,Goodrich_1995}, it is only recently have such hypervariable quasars been observed in great numbers and extended to the high luminosity and high redshift regime \citep[e.g.,][]{Shappee_etal_2014,LaMassa_etal_2015,MacLeod_etal_2016,MacLeod2019,Ruan_etal_2016,Runco_etal_2016,Runnoe_etal_2016,Graham_etal_2017,Rumbaugh_etal_2018,Yang_etal_2018,Graham_etal_2019}. Detailed follow-up observations with optical spectroscopy, polarimetry and mid-infrared imaging of individual objects suggest that the dramatic flux variability is mainly caused by intrinsic variability of the accretion disk emission \citep[e.g.,][]{Denney_etal_2014,LaMassa_etal_2015,Runnoe_etal_2016,Hutsemekers_etal_2017,Sheng_etal_2017,Ross2018,Dexter_etal_2019}, rather than by variable dust obscuration or transient events such as tidal disruption events \citep[e.g.,][]{Merloni_etal_2015}. However, there are a few cases where the dramatic rise of a quasar light curve can be reasonably well explained by microlensing \citep[e.g.,][]{Lawrence_etal_2016,Bruce2017}. Recently, \citet{Rumbaugh_etal_2018} studied the statistical properties of a large sample of extreme variability quasars\footnote{\citet{Rumbaugh_etal_2018} used the term ``extreme variability'' to denote this population of hypervariable quasars and adopted a threshold variability of $\Delta g=1$ mag. This definition is less ambiguous than the term ``changing-look quasars'' that are often used in the literature for these objects, which requires additional criteria based on spectroscopy. \citet{Guo_etal_2020} have shown that the ``changing-look'' in the quasar broad emission lines is a natural consequence of photoionization and the extreme variability from the quasar continuum. } (EVQs) and compared to normal quasars. They found that EVQs have on average lower Eddington ratios compared to normal quasars, providing further support to the intrinsic origin of the extreme variability. 

The large-amplitude intrinsic variability of the accretion disk flux is difficult to explain in the canonical, steady-state, accretion disk model \citep{Shakura_Sunyaev_1973}. The viscous timescale associated with significant accretion rate changes in this model is several orders of magnitude longer than the few-year timescales observed for these EVQs. There have been several attempts to explain the observed dramatic variability on multi-year timescales that invoke various instabilities in the accretion disk region \citep[e.g.,][]{JiangY_etal_2016,Ross2018,Dexter_Begelman_2018}.

In this work, we study the largest sample of EVQs with well-sampled light curves that cover $\sim 16$ years in the observed frame. By studying the long-term light curves of these EVQs we hope to establish an observational consensus of the EVQ phenomenon on decade-long timescales, and shed light on the nature of these large-amplitude variations from the accretion disk of quasars. 

\begin{figure} 
        \centering \includegraphics[width=\columnwidth]{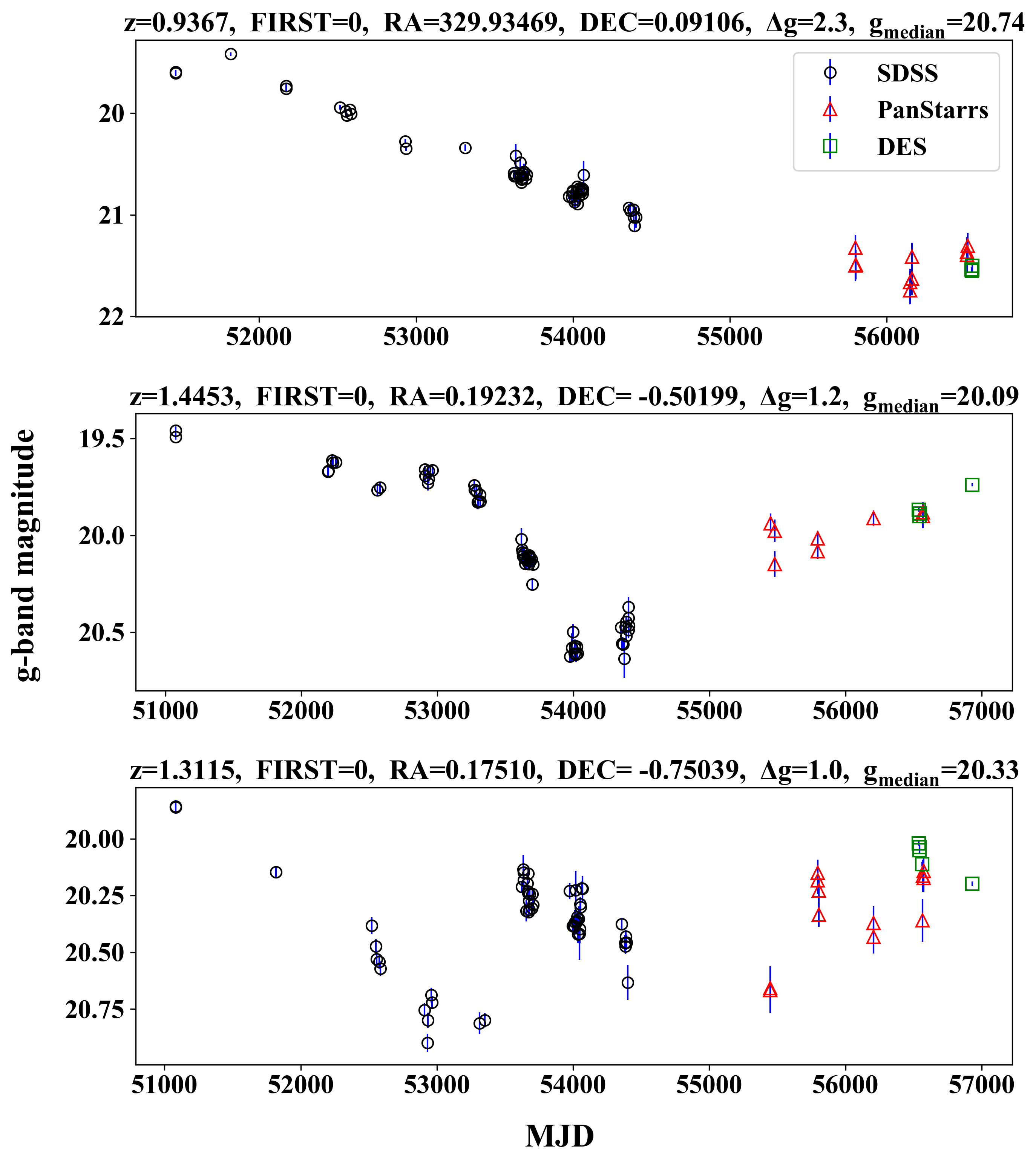}
        \caption{
                \label{fig:examples} 
                Three examples of 16-yr $g$-band light curves of an EVQ from \citet{Rumbaugh_etal_2018}. Different symbols/colors indicate different surveys. The top panel shows an example for a monotonic light curve over the entire period; the middle panel shows an example with a single broad dip in the light curve; and the bottom panel shows an example with a more complex pattern in the 16-yr light curve. All three examples are not detected in the FIRST radio survey \citep[FIRST=0;][]{White_etal_1997}.
        }
\end{figure}

One motivation of our work is the apparent diversity of the long-term light curves among EVQs. Figure \ref{fig:examples} presents examples of 16-yr light curves for several EVQs that have apparently different appearances in their long-term luminosity evolution: some are ``smoother'' in their long-term evolution than the others. It would be interesting to explore if the long-term evolution depends on the physical properties of the quasar such as the BH mass and Eddington ratio. 

The paper is organized as follows. In \S\ref{sec:data} we describe the sample of EVQs and the light curve data. In \S\ref{sec:analysis} we characterise the light curves based on their long-term trend, and study their dependencies on short-term variability as well as other physical properties of the quasar. We discuss the implications of our findings in \S\ref{sec:disc}, and present a sample of 16 EVQs whose long-term extreme variability can be reasonably well explained by a microlensing event (\S\ref{sec:disc_micro}). We summarize our results in \S\ref{sec:con}.

\begin{figure} 
        \centering \includegraphics[width=\columnwidth]{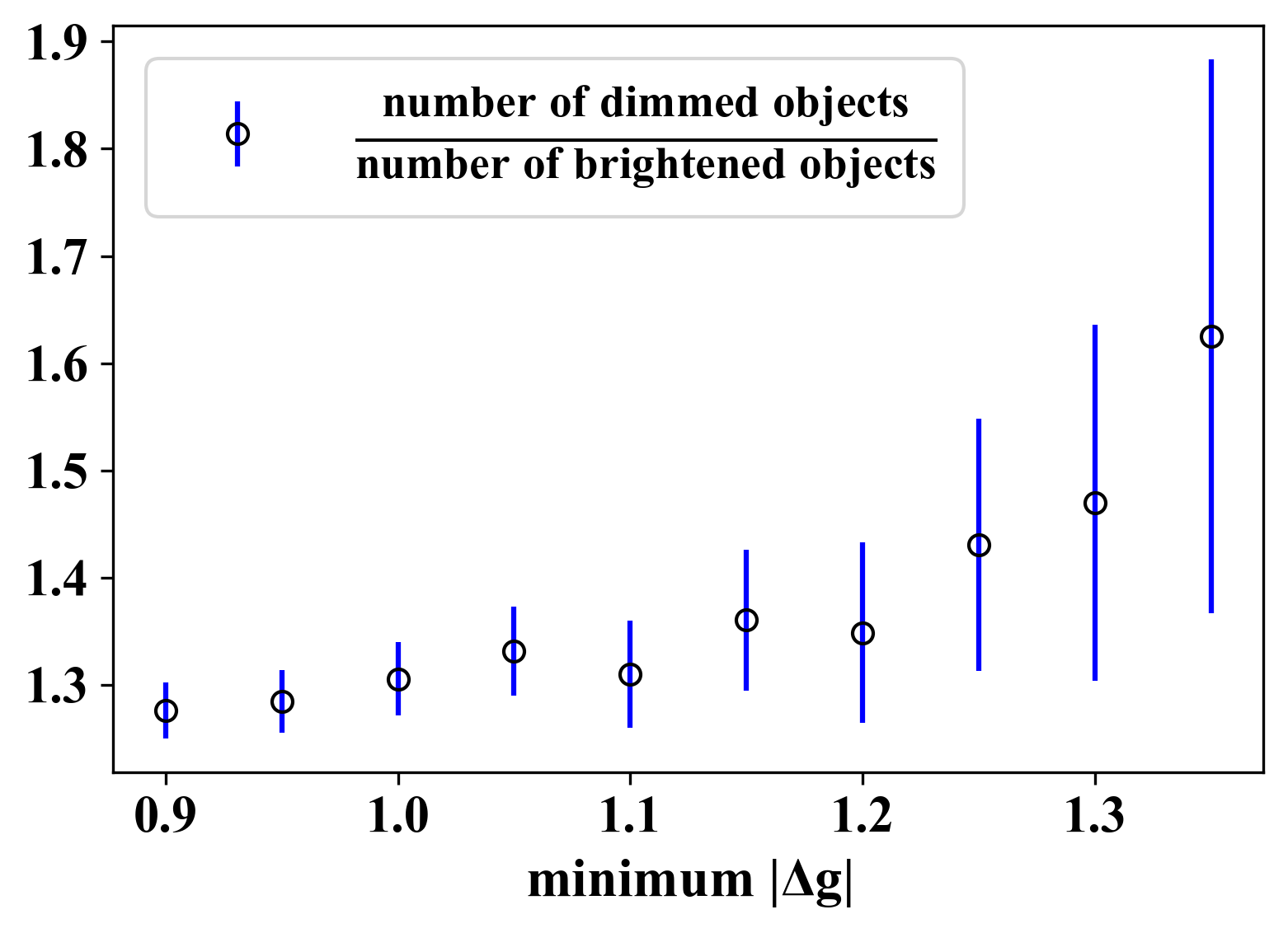}
        \caption{
                \label{fig:bias} 
                Number ratio between dimmed and brightened quasars as a function of the threshold magnitude change from simulations, overplotted with Poisson error bars. There is a strong asymmetry between the numbers of dimmed and brightened quasars due to selection effects (see \S\ref{sec:bias}).
        }
\end{figure}

\begin{figure*} 
        \centering \includegraphics[width=2\columnwidth]{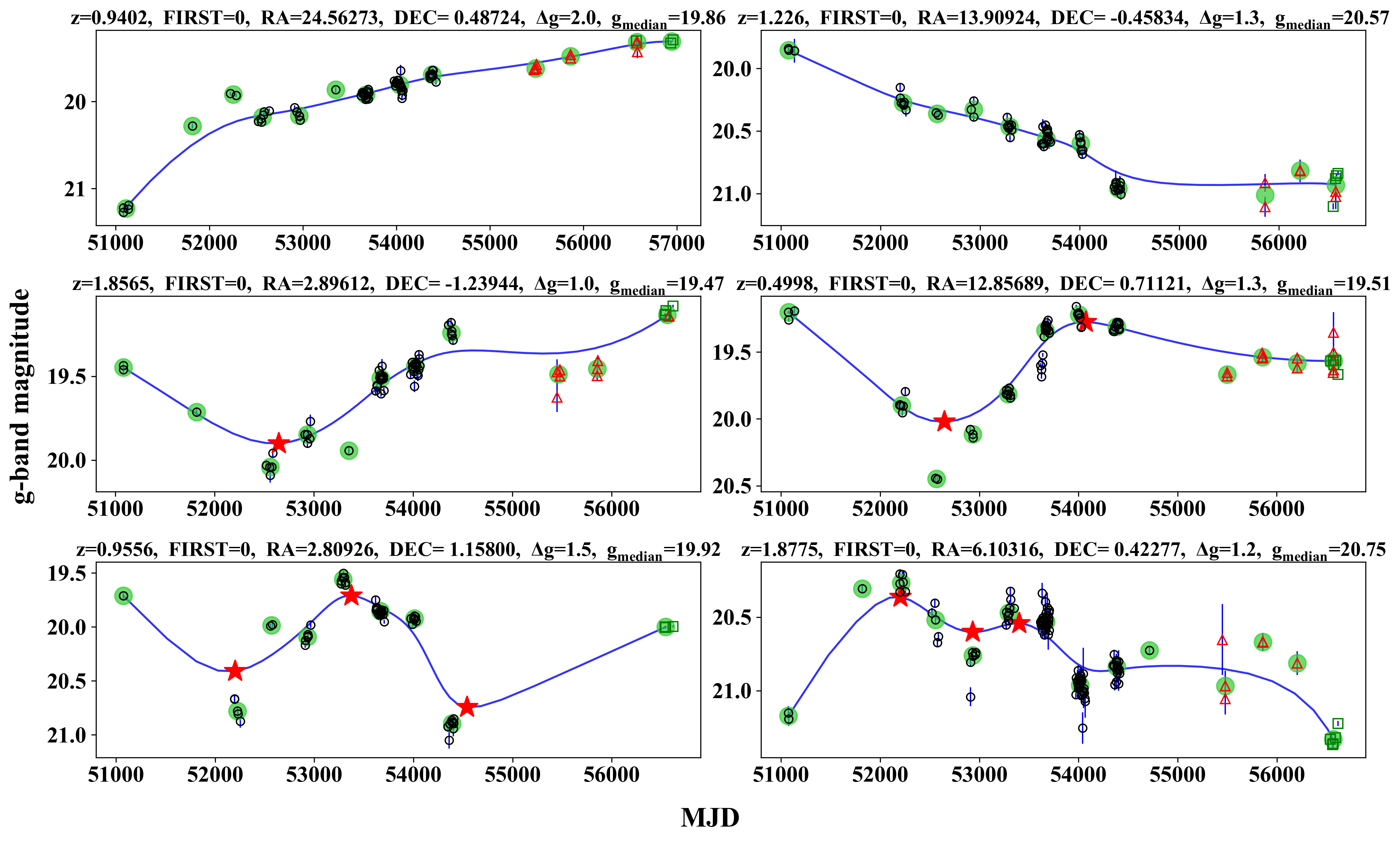}
        \caption{
                \label{fig:example2} 
                Examples of light curves in the three categories we classify. The top, middle and bottom panel each shows two light curves classified as ``monotonic trend'', ``single peak/valley'', and ``complex'' (see \S\ref{sec:long_var}). The solid green points represent the median of data points in each 6-month season. The blue curve is the B\'{e}zier curve (described in \S\ref{sec:long_var}) constrained using the seasonal median points as control points, which is a smoothed representation of the multi-year light curve that we use to classify the long-term trend. The red stars mark the identified inflection points (described in \S\ref{sec:long_var}) on the B\'{e}zier curve.
        }
\end{figure*}

\section{Sample and Data}\label{sec:data}

We use the large sample of $\sim 900$ EVQs compiled in \citet{Rumbaugh_etal_2018} that are covered in the SDSS Stripe 82 region. All of these quasars varied by more than 1 magnitude in $g$ band over the 16-yr baseline. For all quasars we also have spectral measurements based on SDSS optical spectroscopy compiled in \citet{Shen_etal_2011}. Most of these EVQs are radio quiet \citep{Rumbaugh_etal_2018}. The light curve data come from three major optical imaging surveys: SDSS, PanSTARRS1, and the Dark Energy Survey. 

SDSS mapped the sky in five filters ($ugriz_{\rm SDSS}$) using a 2.5 m telescope \citep{Gunn_etal_2006} at the Apache Point Observatory \citep{Abazajian2009}, covering 11,663 deg$^2$ of the sky. SDSS also repeatedly imaged a $120^\circ \times 2.5^\circ$ stripe along the celestial equator centered at zero declination in the Southern Galactic Cap (``Stripe 82") from 1998 to 2007. The observing cadence was increased from 2005 to 2007 for the supernova survey \citep{Frieman2008}. Over the 10 year duration of the program, there are, on average, more than 60 epochs in the Stripe 82 region. 

The Pan-STARRS1 \citep[PS1;][]{Chambers2016} survey mapped three-quarters of the sky in five broadband filters ($grizy_{\rm PS1}$), using a 1.8-m telescope with a 1.4 Gigapixel camera. PS1 data were obtained during $\sim 2010- 2014$, filling the gap between the SDSS and DES surveys. We used the multi-epoch photometry from the detection catalog in the PS1 Data Release 2 (DR2). For quasars, the magnitude offset between the SDSS and PS1 (DES) $g$ bands is negligible, between $-0.053$ ($-$0.065) and 0.005 (0.008) mag at redshift $z<2$ \citep{Yang_etal_2018}. 


The Dark Energy Survey (DES) observed 5000 deg$^2$ of the sky in five filters ($grizY_{\rm DES}$), using a wide-field camera (DECam) on the 4-m Blanco Telescope \citep{Flaugher2015}. The 10$\sigma$ single-epoch PSF magnitude limit in the five $grizY$ bands are 23.57, 23.34, 22.78, 22.10, and 20.69, respectively \citep{Abbott2018}. We use the 3-yr DES light curves from Aug 2013 through Feb 2016 \citep{Diehl_etal_2016}, as compiled by \citet{Rumbaugh_etal_2018}. 

Fig.~\ref{fig:examples} shows several example light curves of our sample, in the order of increasing complexity of the long-term trend. In addition to the extreme variability (more than one magnitude in $g$), these examples demonstrate the diversity of the long-term trend of the light curves of EVQs. 

\section{Light curve characterization}\label{sec:analysis}

The optical variability of quasars is stochastic in general. In this work we distinguish the variability on short timescales and on long timescales using simple metrics of variability computed from the light curve. For ``short-term'' variability we refer to variability on $<1$ yr timescales. This division between short- and long-term obviously is arbitrary, and is primarily motivated by the seasonal gaps in these light curves. Nevertheless, it allows us to explore the different behaviors of variability occurring on short (seasonal) and multi-year timescales. 

The short-term variability is estimated as the RMS magnitude using all light curve points in each season, then averaged (taking the median) over multiple seasons. In computing the RMS magnitude, we subtract the median magnitude uncertainty in each season in quadrature to remove the contribution from measurement uncertainties. We set the RMS magnitude to zero in cases where the magnitude uncertainty is too large to constrain the intrinsic RMS variability. 

For long-term variability, we are more concerned with the shape of the evolution rather than the RMS magnitude, since by selection these EVQs have large variability on multi-year timescales already. Because there is no a priori knowledge on the classification of long-term light curve trends, we rely on empirical approaches with a combination of automated classification and manual inspection. \S\ref{sec:long_var} describes our best effort in classifying the long-term trends. 

\subsection{A selection bias in EVQs}\label{sec:bias}

Before we characterize the light curve patterns, we quantify a selection bias in the search of hypervariable quasars from multiple surveys. Consider a sample of quasars confirmed in a prior survey (e.g., SDSS), when we observe them in a later survey (e.g., DES), we often find more dimmed quasars than brightened ones \citep[e.g.,][]{Rumbaugh_etal_2018}. This asymmetry is negligible for small magnitude changes but becomes significant for large magnitude changes. \citet{Rumbaugh_etal_2018} pointed out that this is due to the flux-limit nature of the surveys, quasar variability and a steep quasar luminosity function (LF), which in combination leads to a selection bias. This bias is more severe for larger variability threshold in defining the variable sample, and therefore for the selection of EVQs the asymmetry between dimmed and brightened objects is substantial. We now investigate this statistical bias with simulations. 

We base our simulations at $z = 1.3$, the median redshift of our EVQ sample. Using the \citet{Hopkins_etal_2007} quasar LF, we simulate a population of quasars down to very faint magnitude ($g\sim$ 22). We then simulate the light curves of these individual objects using the Damped Random Walk model \citep[e.g.,][]{Kelly_etal_2009,Kozlowski_etal_2010,MacLeod_etal_2010} that provides a reasonably good description of the stochastic optical quasar variability. We set $SF_{\infty} = 0.2$ mag and rest-frame damping timescale $\tau=200$ days based on results in \citet{MacLeod_etal_2010} for Stripe 82 quasars. At a prior epoch ($t=0$) we select a flux-limited sample ($g<21$) to mimic the parent quasar sample from SDSS. At a later epoch ($\Delta t=10$ yr in the rest frame of the quasar), we compile the evolved magnitudes for this flux-limited sample. Fig. \ref{fig:bias} displays the ratio between dimmed and brightened quasars as a function of the threshold magnitude change over the entire light curve. There is a strong asymmetry in the numbers of dimmed and brightened quasars, which increases with the threshold magnitude change. Out of their 977 EVQs with $|\Delta g|>1$, \citet{Rumbaugh_etal_2018} observed 372 brightened and 605 dimmed between the two extreme states, which corresponds to a ratio of $\sim$1.6 (dimmed/brightened). In our simulation, this ratio is $\sim 1.3$ for the same $|\Delta g|$ threshold as shown in Fig.~\ref{fig:bias}. The exact amount of this asymmetry depends sensitively on the slope of the quasar LF, the redshift range and the flux limit of the sample. Therefore the simulation predictions do not match exactly with the observed number statistics in the \citet{Rumbaugh_etal_2018} EVQ sample, which covers a broad redshift range and is not exactly flux-limited. Nevertheless, the bias and the trend with the threshold $|\Delta g|$ are qualitatively reproduced in this simulation.

\begin{figure} 
        \centering \includegraphics[width=\columnwidth]{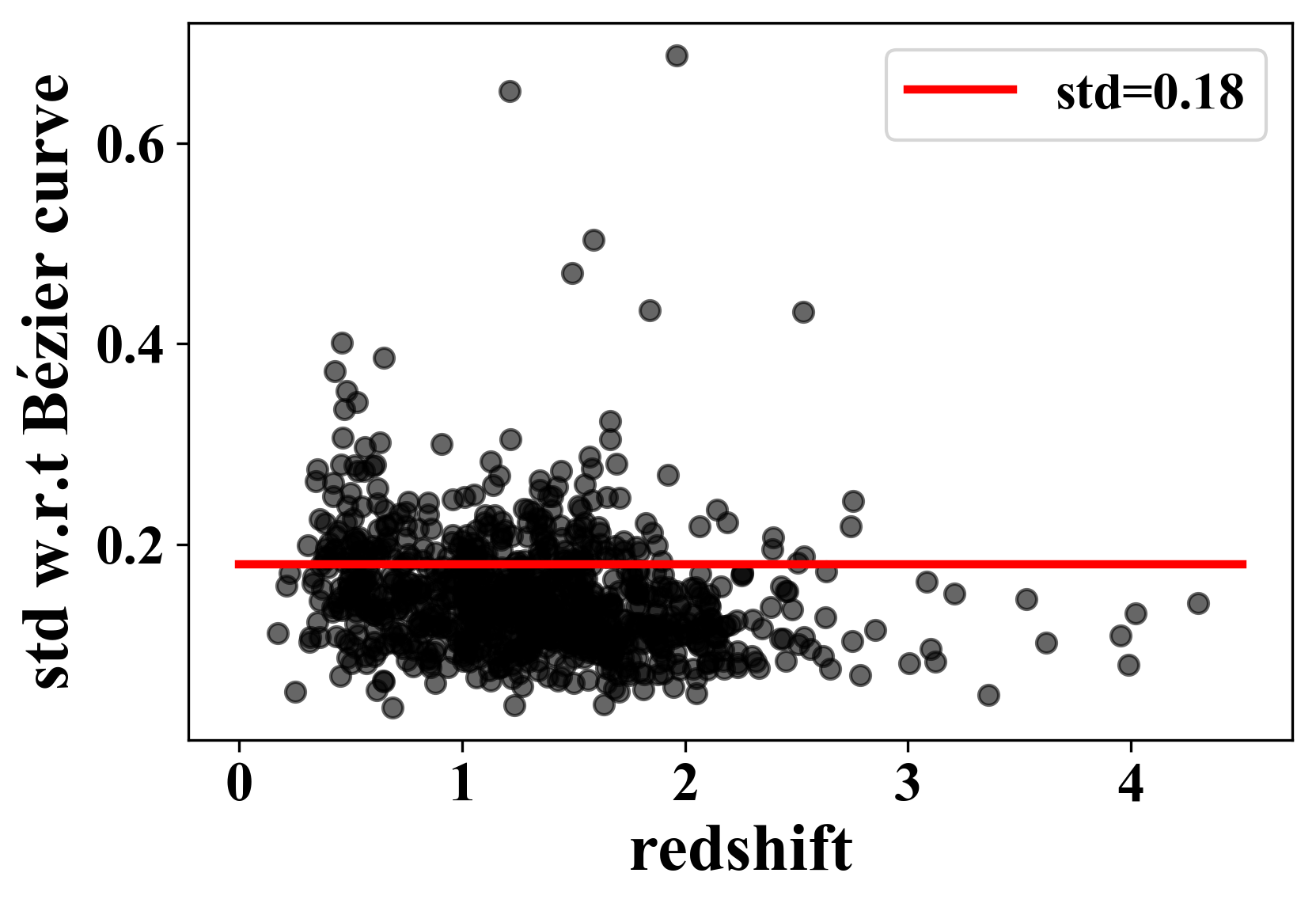}
        \caption{
                \label{fig:std} 
       Redshift versus standard deviation with respect to the B\'{e}zier curve (the smoothed representation of the multi-year light curve trend). 
        }
\end{figure}

\begin{figure} 
        \centering \includegraphics[width=\columnwidth]{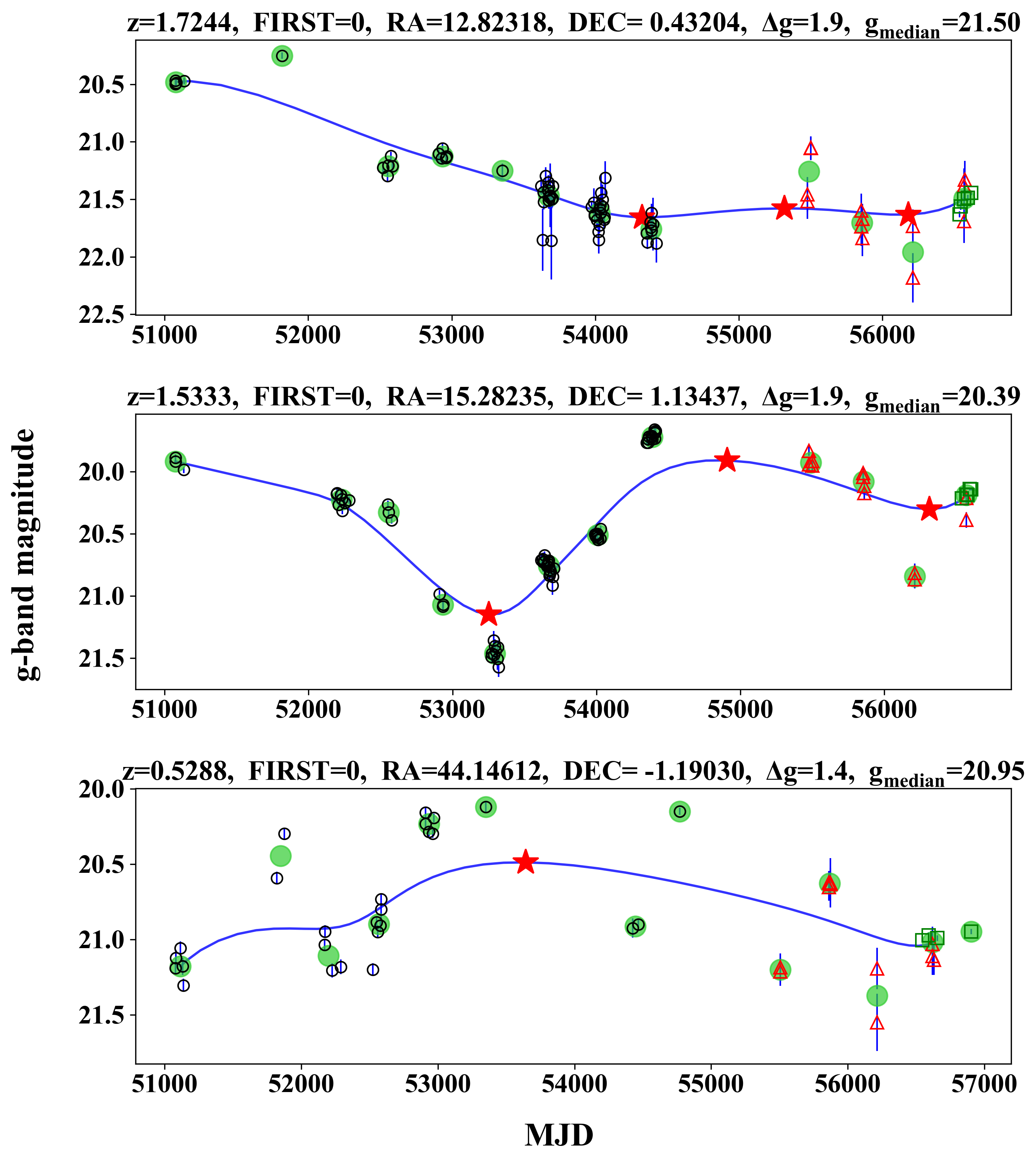}
        \caption{
                \label{fig:examples-corr} 
              Examples of objects which we manually correct for classification. The first panel shows an object that is originally classified as ``complex'' but corrected to ``monotonic trend''. The middle panel shows an object that is originally classified as ``complex'' but corrected to ``peak/valley''. The bottom panel shows and object that is originally classified as ``peak/valley'' but corrected to ``complex''. The color scheme and symbols in this plot are the same as in Fig.~\ref{fig:example2}.  
        }
\end{figure}

\begin{figure} 
        \centering \includegraphics[width=\columnwidth]{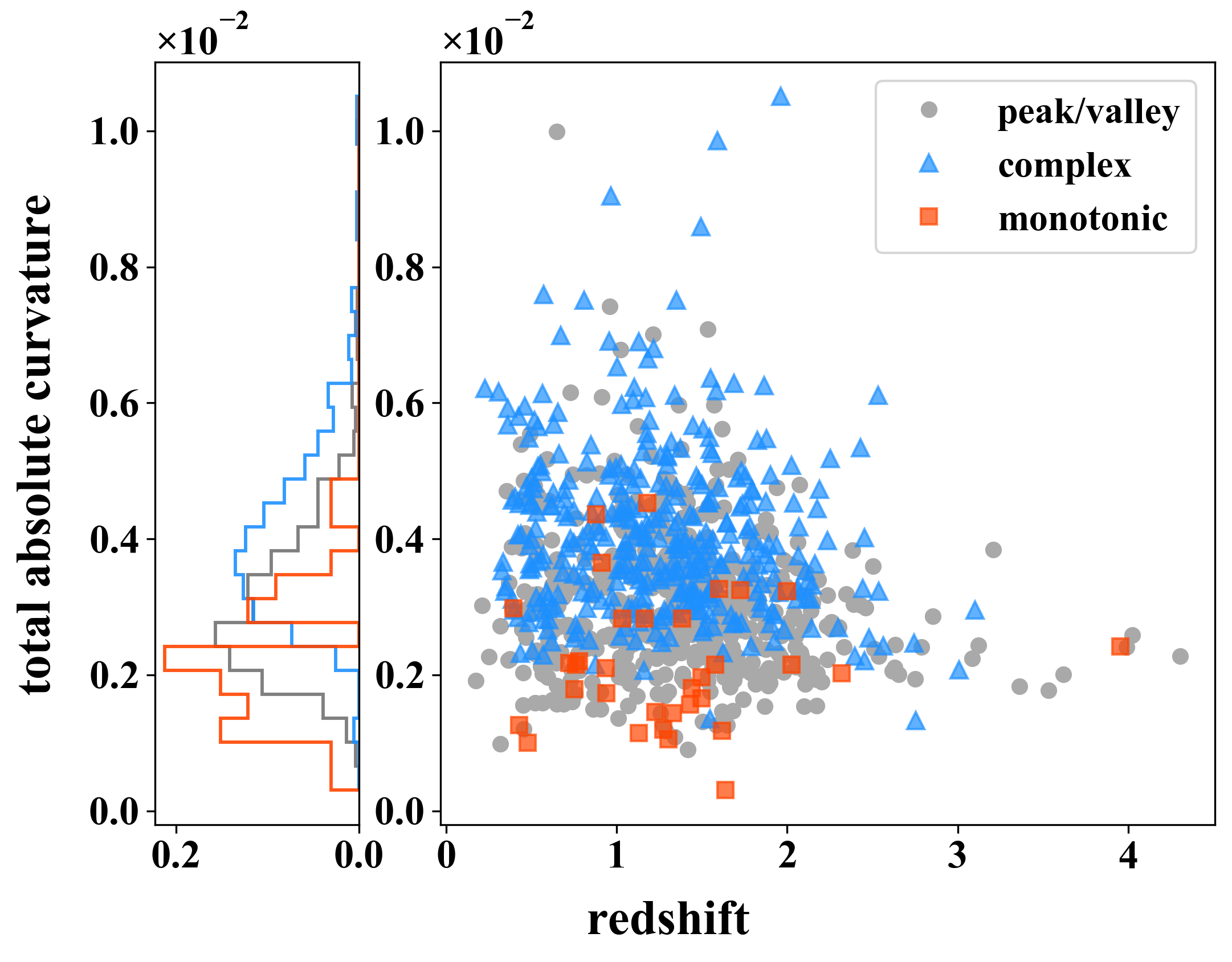}
        \caption{
                \label{fig:z-cur} 
      Distribution of EVQs in the three categories (mono, single peak/valley, complex; see \S\ref{sec:long_var}) in the total absolute curvature versus redshift plane. The total absolute curvature increases from mono to complex, as expected from the definitions of these categories.  
        }
\end{figure}

\begin{figure*} 
        \centering \includegraphics[width=2\columnwidth]{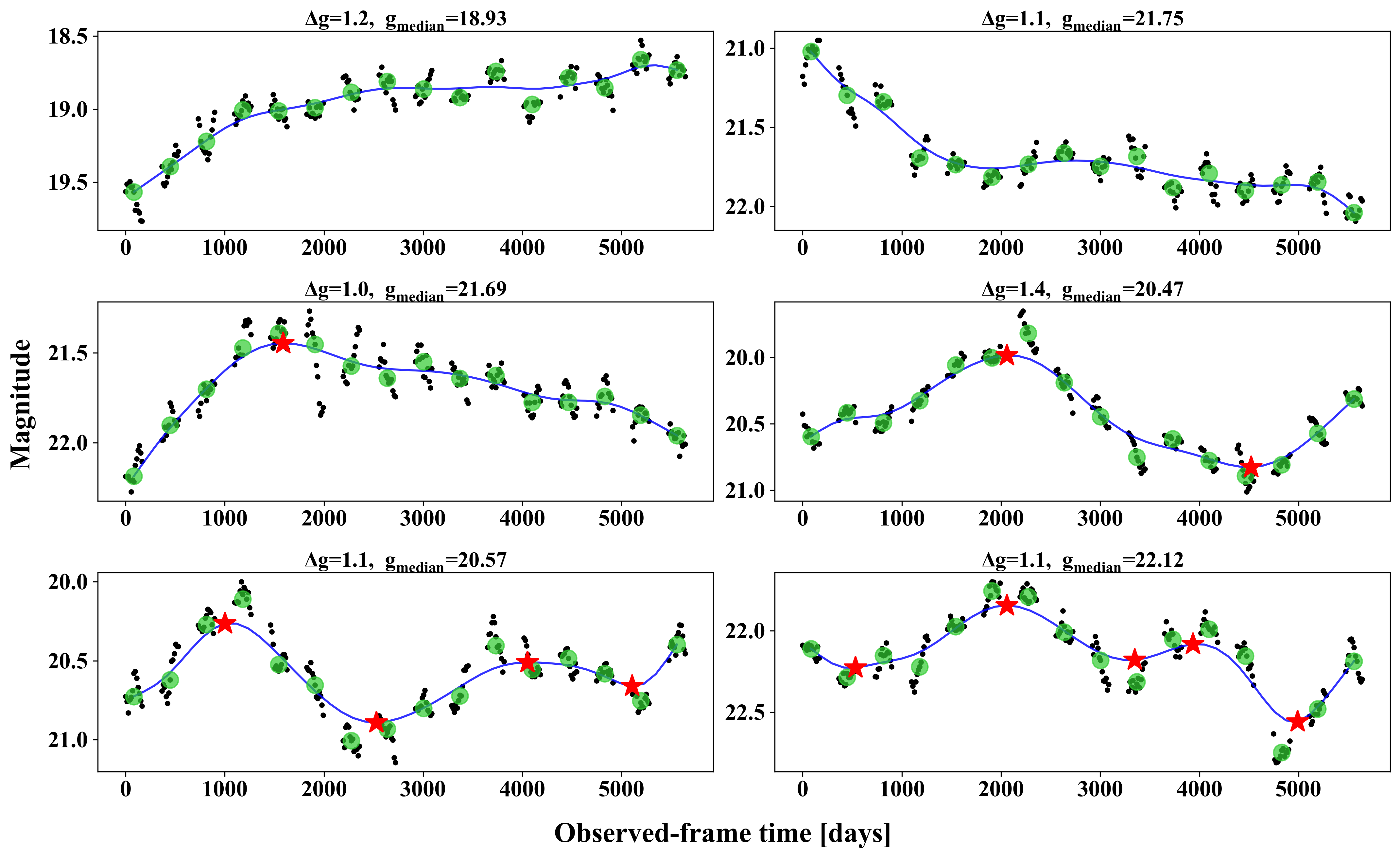}
        \caption{
                \label{fig:example-drw} 
                Examples of simulated DRW light curves, which we classify into the three categories as for the real data. The top, middle and bottom panel each shows two light curves which we classify as ``monotonic trend'', ``single peak/valley'', and ``complex'' (see \S\ref{sec:long_var}, \S\ref{sec:long_var_drw}). The color scheme and symbols in this plot are the same as in Fig.~\ref{fig:example2}.
        }
\end{figure*}

\subsection{Long-term (multi-year) variability}\label{sec:long_var}
When classifying the long-term trend of the light curves, we want to identify the dominant long-term trend over the entire $\sim$16-yr baseline. The original data points in the light curves are clustered in seasonal groups and scattered due to short-term variations. This motivated us to create a smoothed mathematical representation of the light curve using the B\'{e}zier curve so that long-term (e.g., over multiple seasons) features in the light curve are easier to identify than using the original light curve. Given a set of points, the B\'{e}zier curve smoothly follows the locations of the points without adding extra features between given points, and thus provides an efficient representation of the distribution of given points. It is also computationally simple to generate. The choice of the B\'{e}zier curve is purely empirical, and we assign no physical meaning to the curve.

We first take the median of data points in each season (shown as solid green points in Fig. \ref{fig:example2}), and then use these seasonal-median points as control points to constrain a smooth B\'{e}zier curve. A B\'{e}zier curve is a parametric curve that uses the Bernstein polynomials as a basis \citep{ComputerAidedDesignBook,mortenson1999mathematics}:
\begin{equation}
    \mathbf{p}(u) = \sum_{i=0}^{n} \mathbf{p}_i B_{i,n}(u),\ \ 0 \leq u \leq 1,
\end{equation} 
where $\mathbf{p}_{i...n}$ are the control points, and $B_{i,n}(u)$ are the Bernstein polynomials
\begin{equation}
    B_{i,n}(u) = \binom{n}{i} u^i (1-u)^{n-i}.
\end{equation} 
We use the \textit{de Casteljau algorithm} \citep{ComputerAidedDesignBook} to evaluate the B\'{e}zier curve, which defines the curve recursively
\begin{equation}
   \mathbf{b}_i^k(u) = (1-u) \mathbf{b}_{i-1}^{k-1} + u \mathbf{b}_{i}^{k-1},\ \ k=1,2,...,n,\ \ i=k,...,n
\end{equation} 
where $\mathbf{b}_i^0$ is the $i$th control point, and the value of the curve at parameter value u is $\mathbf{b}_n^n$.

To account for the different numbers of epochs in each season in the construction of the B\'{e}zier curve, we up-weight the control point (seasonal median) by two if there are more than three epochs, and by three (four) if there are more than five (seven) epochs in that season. In addition, since he B\'{e}zier curve is designed to pass through the boundary control points, if there is only one epoch in the last season of the light curve, that seasonal median (the data point itself) is not used as a control point. These empirical recipes are implemented for the B\'{e}zier curve to best represent the overall long-term trend of the light curve. The B\'{e}zier curve then roughly represents the long-term smoothed trend of the light curve, which we use to further classify the light curves. 

To quantify the complexity of the B\'{e}zier curve, we count the number of inflection points (local extrema) of the curve. In order to capture only prominent features in the curve, we calculate the topographic prominence of each identified point of local extrema, and only denote that point as an inflection point when its topographic prominence is greater than 0.05 (mag). Topographic prominence measures the height of the peak with respect to the lowest contour line enclosing the peak but containing no higher peak within it. This quantity characterizes how much a peak stands out from the surrounding baseline. Again this definition of inflection points here is empirical and merely serves the purpose of quantifying prominent features in the representations of light curves.

With the identified inflection points, we classify each light curve into three categories: (1) monotonic trends (mono): zero inflection point; (2) single broad peak and/or dip (peak/valley): one or two inflection point(s); (3) more complex patterns (complex): more than two inflection points.

There are cases where the B\'{e}zier curve cannot well represent the running trend of the light curve, e.g., due to overly scattered distribution of control points, or skew of the curve shape by one or two particular control points.
Thus the above scheme might misclassify the light curve. To correct for these cases, we calculate the vertical standard deviation of the control points with respect to the corresponding B\'{e}zier curve. Fig.~\ref{fig:std} displays the result for all objects, and the majority of them have reasonably small standard deviations. We visually inspect the light curves that have standard deviations larger than 0.18 (mag) and correct for the classification if necessary. The numbers of manually corrected objects in category mono, peak/valley, and complex are 4, 11, 50, respectively. An example of manually corrected objects in each category is shown in Fig.~\ref{fig:examples-corr}. We summarize the statistics of each of the three categories in Table \ref{table: category}.

To further demonstrate that objects in the three categories as classified above have different long-term light curve smoothness, we calculate the total absolute curvature
\begin{equation}
    \int |\kappa (s) |ds,
\end{equation}
where $s$ is the arc length parameter and $\kappa$ is the curvature, of the B\'{e}zier curve for each EVQ. The results are plotted against redshift in 
Fig. \ref{fig:z-cur}. Although there is substantial overlap among the three categories, the total absolute curvature of the B\'{e}zier curve representing the light curve on average increases from category mono to category complex. The median total absolute curvature for each category is presented in Table \ref{table: category}.

\subsection{Long-term (multi-year) trends of simulated light curves}\label{sec:long_var_drw}
In \S\ref{sec:bias}, we generated 36784 simulated light curves at redshift $z=1.3$ with the Damped Random Walk (DRW) model to test a selection bias in EVQs. These light curves span a baseline of 10 years in the source rest frame and have one data point per day. Here we first shift the time series to the observed frame and truncate the baseline at 16 years, then downsample the light curves to mimic the observational cadence in real data. In real data, the exact number of data points per season varies from one object to another, and occasionally there are gaps of longer than one year. Here we adopt a simple scheme such that the cadence is uniform for each season, resulting in more uniformly-sampled mock light curves and slightly more individual epochs than the real light curves. However, since we will measure the long-term trends using seasonally binned data, such simplifications in the mock light curve generation should not impact our results much. We then identify the subset of mock EVQs following the same definition ($\Delta g>1$ over the downsampled light curves), and 4812 mock quasars are selected as EVQs. We apply the same method discussed in the previous section to these simulated light curves and classify them into the same three categories of long-term trends. We find that the fractions of objects for categories mono, peak/valley, and complex are {0.166}\%, {25.5}\%, and {74.3}\%, respectively. Due to the large number of simulated light curves, we do not manually correct for these categories. Calculating the standard deviation of the control points with respect to the corresponding B\'{e}zier curve shows all objects have standard deviations less than 0.18, which also indicates that the smoothed representations of the light curves are mostly reasonable and manual corrections would not change the result substantially. In Fig.~\ref{fig:example-drw} we show two examples of the DRW light curves in each category.

The DRW prescription of quasar variability is simplified and may not well describe the variability on multi-year timescales \citep[e.g.,][]{Guo_etal_2017}, but our light curves are of insufficient quality to test more complicated variability models. However, as shown in \citet{MacLeod_etal_2010}, there is a broad correlation between $SF_{\infty}$ and $\tau$ for SDSS quasars. For EVQs, they tend to have on average larger values of $SF_{\infty}$ and $\tau$ than the rest of the quasars \citep[][]{Rumbaugh_etal_2018}, although some of the highest measured $\tau$ values may suffer from the limited baseline of the light curves \citep[e.g.,][]{Kozlowski_2017}. Larger values of $\tau$ will preserve the $f^{-2}$ red-noise power spectrum density of quasar variability to longer timescales, which will lead to ``smoother'' long-term trends.   

To test the effects of DRW parameters on the resulting simulated light curves, we generated another set of 36784 mock light curves following the same approach above, but with different DRW parameters $SF_{\infty} = 0.4$ mag and $\tau=600$ days, which better represent those of the EVQs \citep[][]{Rumbaugh_etal_2018}. Applying the same classification scheme on the downsampled mock light curves for the subset of EVQs yields $\sim 0.382\%$, $\sim 32.9\%$, $\sim 66.7\%$ of objects in categories mono, peak/valley, complex, respectively. Indeed using larger values of $\tau$ leads to higher fractions of more ``smoother'' light curves on multi-year timescales, as expected. 

Compared with the observed fractions of the three categories, $\sim 4\%$, $\sim 56\%$ and $\sim 40\%$, the simulated light curves slightly under-produce the mono class and the peak/valley class while over-produce the complex class. This may imply that not all EVQ light curves are purely produced by stochastic processes; alternatively, at least not all EVQ light curves can be well described by the DRW model over the multi-year baseline. Nevertheless, the DRW prescription is able to produce many relatively smooth long-term light curves that we can classify as mono or single peak/valley, indicating that such light curves are consistent with being produced by stochastic processes.

\begin{figure} 
        \centering \includegraphics[width=\columnwidth]{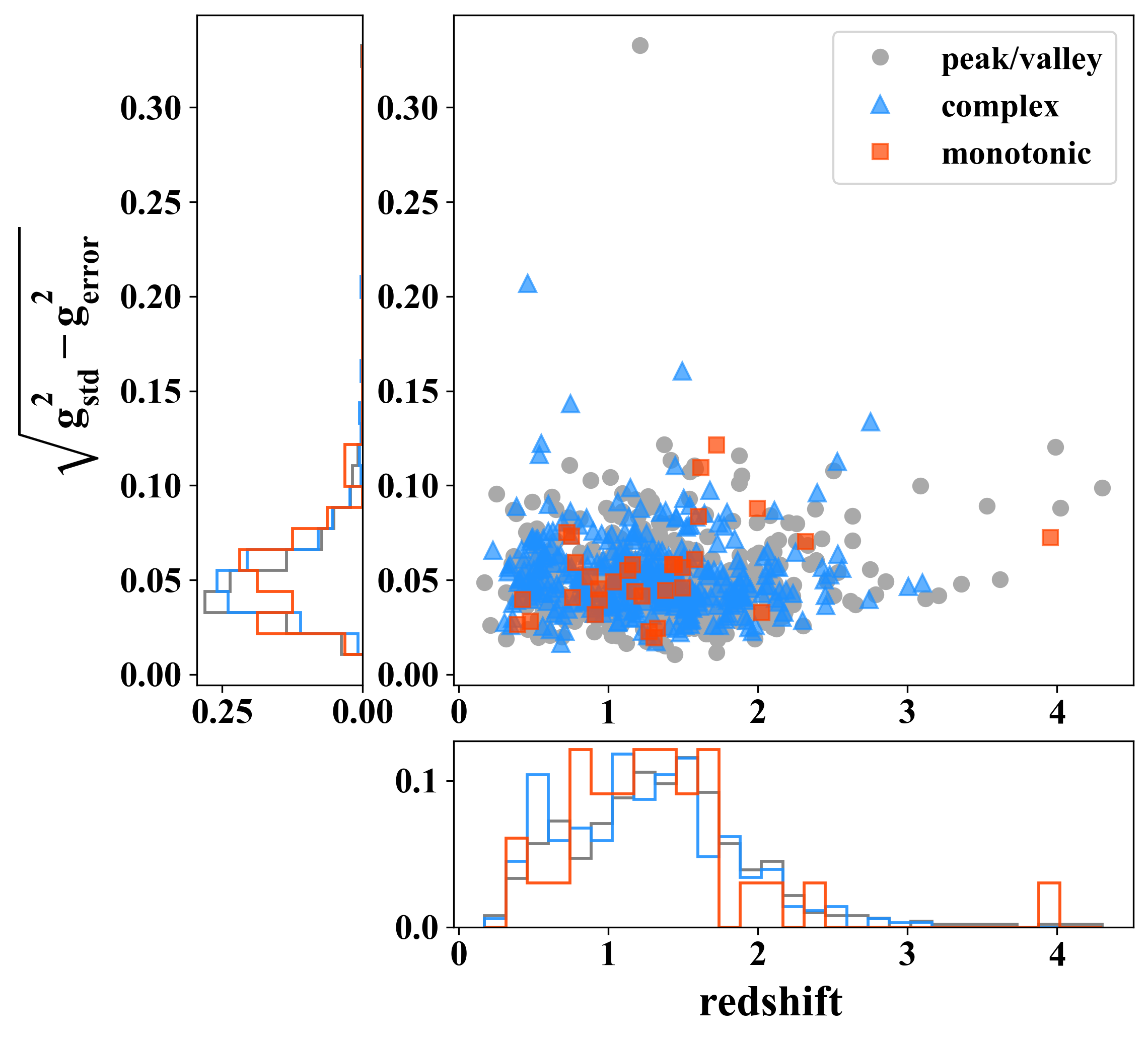}
        \caption{
                \label{fig:z-vari} 
                Distributions of EVQs in the three categories (mono, single peak/valley, complex; see \S\ref{sec:long_var}) in the redshift versus short-term (seasonal) variability plane. There is no significant difference among the three categories. 
        }
\end{figure}

\begin{figure} 
        \centering \includegraphics[width=\columnwidth]{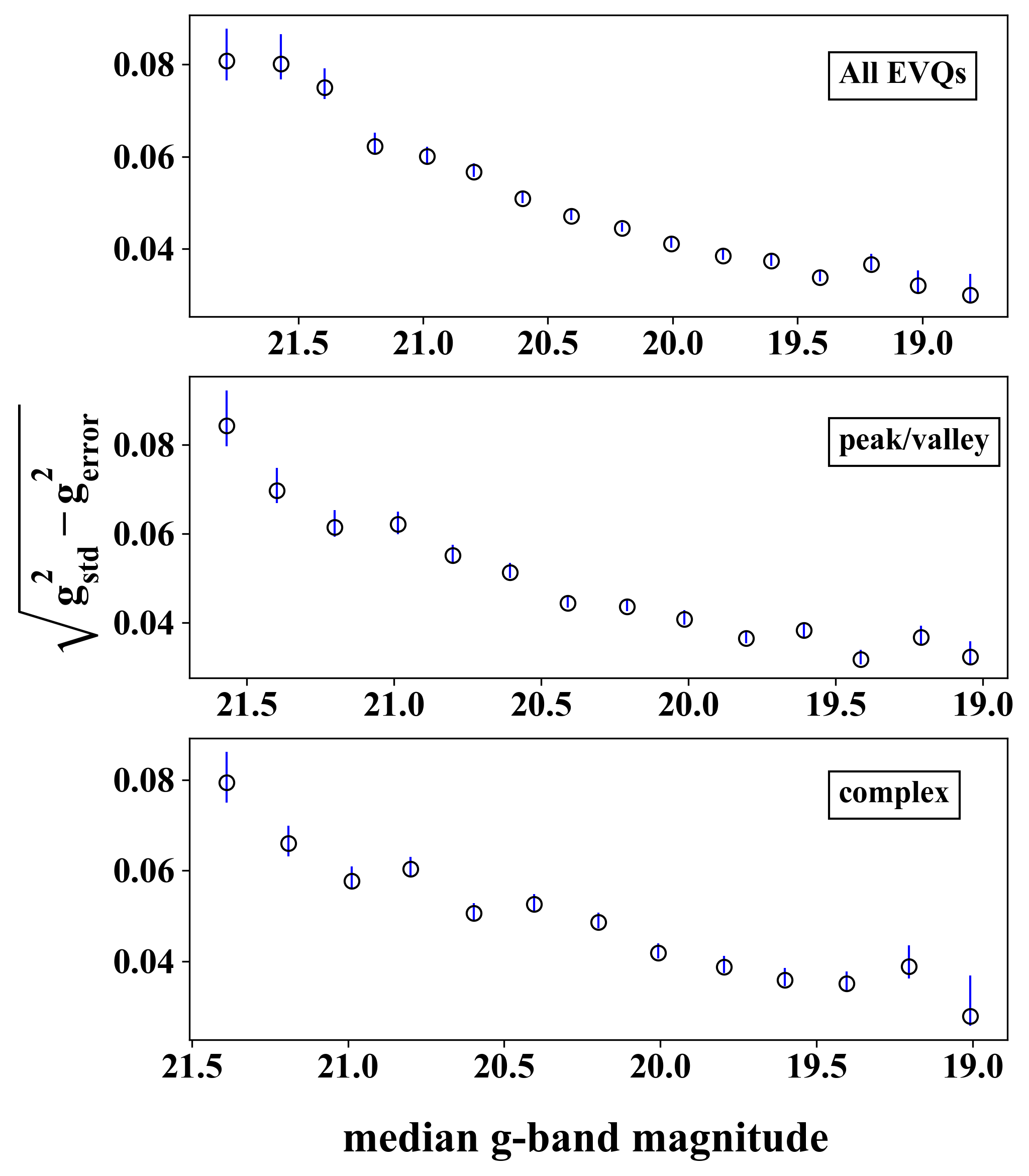}
        \caption{
                \label{fig:rsm-mean1} 
                The ensemble relation between the short-term (seasonal) variability and the seasonal-median flux for the EVQ sample. In each seasonal-median magnitude bin, we plot the median short-term variability from all objects (open circles) and the uncertainty in the median (the 16-84\% range divided by $\sqrt{N}$, with $N$ being the number of objects in the bin). There is a general trend that the average short-term variability in magnitude decreases when the seasonal-median flux increases. The top panel shows the result for all EVQs, and the middle and bottom panels show that for the single peak/valley and complex categories. The monotonic trend category does not have enough objects to measure this ensemble relation.
        }
\end{figure}

\begin{figure} 
	\centering \includegraphics[width=\columnwidth]{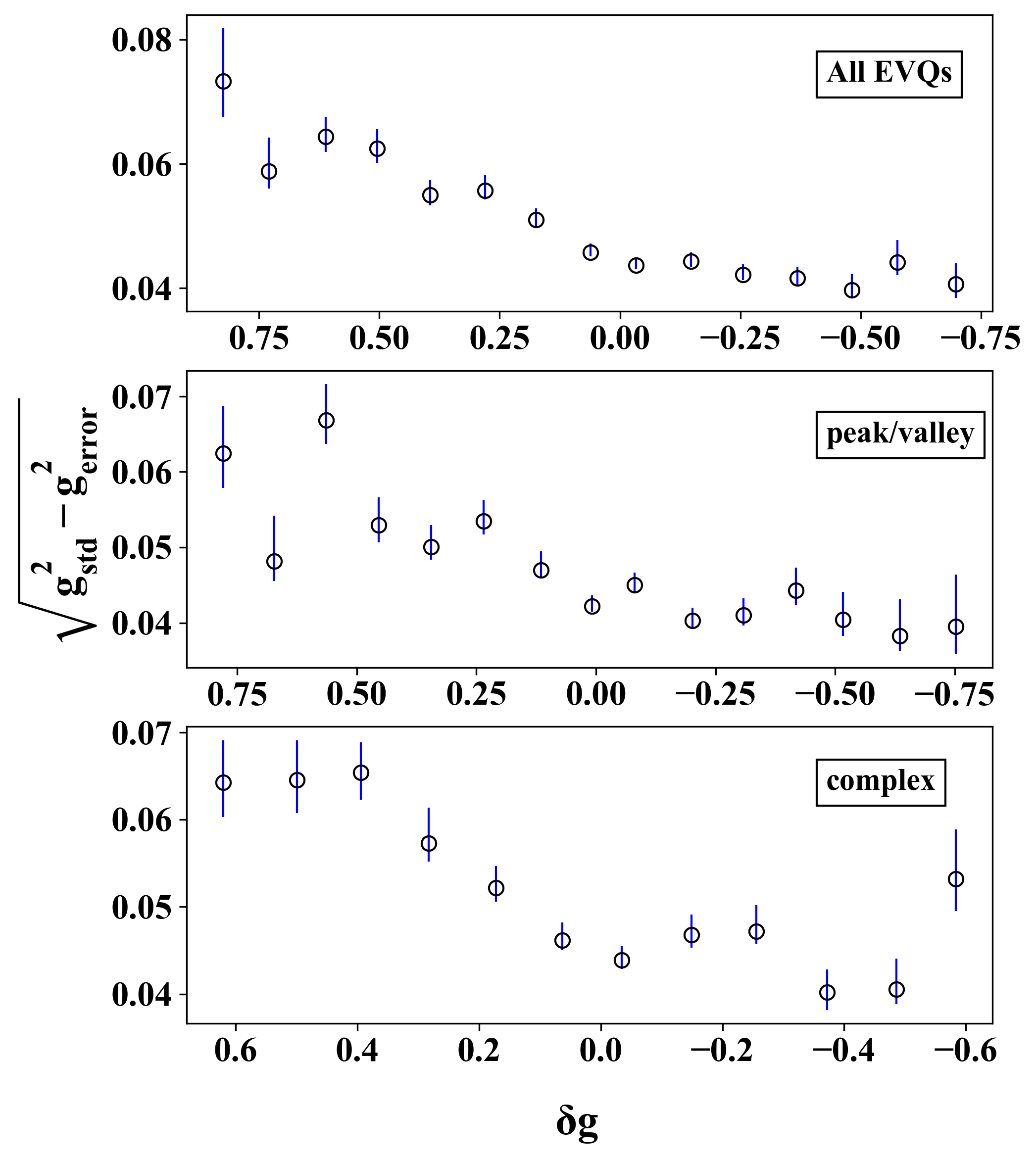}
	\caption{
		\label{fig:rsm-mean_dg2} 
		The ensemble relation between the short-term (seasonal) variability and the difference between the seasonal median magnitude and the median magnitude of the entire light curve. Symbols and notations are the same as in Fig.~\ref{fig:rsm-mean1}.
	}
\end{figure}

\begin{table*}
\begin{center}
\caption{Statistics of three EVQ categories as described in \S\ref{sec:long_var}. The radio-loud fraction is estimated assuming all FIRST-detected quasars are radio-loud, and is not corrected for selection completeness in FIRST given that the EVQ sample includes quasars as faint as $g\approx 22$ \citep{Rumbaugh_etal_2018}.}
\label{table: category}
\begin{tabular}{c p{1.2cm} p{1.3cm} p{1.3cm} p{1.3cm} p{1.3cm} p{1.5cm} p{1.3cm} } \hline
Category & Object number & Number of inflection points & Median redshift & Median log$L_{\rm bol}$ [erg/s] & Median log virial BH mass [M$_{\sun}$] & Median total absolute curvature [10$^{-3}$] & Radio loud fraction \\ \hline
monotonic & 33 & 0 & 1.273 & 45.84 & 9.00 & 2.1 & 6.1\%\\ 
single broad peaks and dips & 510 & 1 or 2 & 1.347 & 45.87 & 8.89 & 2.9 & 9.2\%\\
complex & 355 & >2 & 1.249 & 45.78 & 8.86 & 3.9 & 9.0\%\\
\hline
\end{tabular}
\end{center}
\end{table*}

\subsection{The rms-mean flux relation}\label{sec:meanrms}

Fig.\ \ref{fig:z-vari} displays the distribution of objects in the redshift versus short-term variability plane, for the three categories of long-term light curve trends. There is no significant difference in the short-term RMS magnitude among different categories. 

The EVQs in our sample span a large dynamic range in flux over the 16-yr light curves. The long-term evolution in most cases is smooth enough such that we can evaluate the short-term variability at different mean luminosity state in each season, and derive the rms-mean flux relation for EVQs. However, for individual objects, the statistics is still poor given the limited number of data points and the uncertainties of the flux measurements. To boost the signal-to-noise ratio of the rms-mean flux relation, we co-add the results from individual objects in the g-band seasonal-median magnitude bin. The resulting ensemble rms-mean flux relations are shown in Fig.~\ref{fig:rsm-mean1}. When creating this ensemble relation, we keep track of the fraction of objects in each magnitude bin that have unconstrained short-term variability due to measurement uncertainties. There is a mild trend that this fraction of immeasurable objects increases from $\sim 40\%$ in the brightest seasonal-median g-band magnitude bin to $\sim 55\%$ in the faintest magnitude bin. However, this small difference will not induce a severe bias in the measured short-term variability in different magnitude bins. 

Fig.~\ref{fig:rsm-mean1} suggests that the short-term variability in magnitude decreases when seasonal-average (median) flux increases. This result differs from X-ray variability studies \citep[e.g.,][]{Gaskell2004,Uttley_etal_2005}, where there is a linear relation between the short-term variability (flickering) and the mean flux, corresponding to a constant fractional rms variability (i.e., magnitude) as the mean flux increases. 

Instead, the observed optical rms-mean flux relation is consistent with the observed anti-correlation between the ensemble variability magnitude and the luminosity of quasars \citep{VandenBerk_etal_2004}. However, our result is the first measurement of the intrinsic rms-mean flux relation that utilizes the large dynamic range in flux changes over long baselines in individual quasars. With continued photometric monitoring of these EVQs to extend the baseline it may eventually become possible to measure the optical rms-mean flux relation for individual objects. 

One potential selection bias is that quasars with lower average luminosities also tend to have larger seasonal rms, which itself may be a manifestation of the intrinsic optical rms-mean flux relation implied above. By including all quasars in the ensemble rms-mean relation, the dynamical range in flux is not only contributed by individual quasar variability, but also the dispersion in the average quasar fluxes, which may lead to an artificial trend as the one observed. To limit the impact of sample dispersion in the average flux, we create a different stack, in which we bin the seasonal-rms magnitude of individual EVQs by their $\delta g\equiv g_{\rm med,season}-g_{\rm med,full}$, the difference between the seasonal-median and the median magnitude over the entire light curve for each quasar. Fig.~\ref{fig:rsm-mean_dg2} shows that the same rms-mean flux relation is roughly preserved.   

The lack of a linear relation between the rms flux and the seasonal-averaged flux (not magnitudes) in the optical suggests different variability mechanisms in the optical for the short-term flickering and long-term evolution of the accretion disk emission. It is possible that the short-term flickering of the optical emission is driven by the accretion disk reprocessing of variable emission from the innermost part of the accretion flow (e.g., the X-ray corona), while the long-term optical variability is driven by instabilities in the accretion disk that are further out and on longer timescales. The variability in the X-ray corona emission, which drives the short-term optical variability (seasonal-rms) by reprocessing, can be largely independent on the long-term evolution of the accretion disk that drives the observed long-term optical variability (changes in seasonal-average). This disconnect between the mechanisms for the short-term and long-term optical variability naturally breaks the observed linear rms-mean relation in X-ray variability. This scenario is supported by observations of correlated UV and X-ray variations on days timescales and the general lack of such correlations on timescales of years \citep[e.g.,][and references therein]{Ulrich_etal_1997}.

\begin{figure*} 
        \centering \includegraphics[width=2\columnwidth]{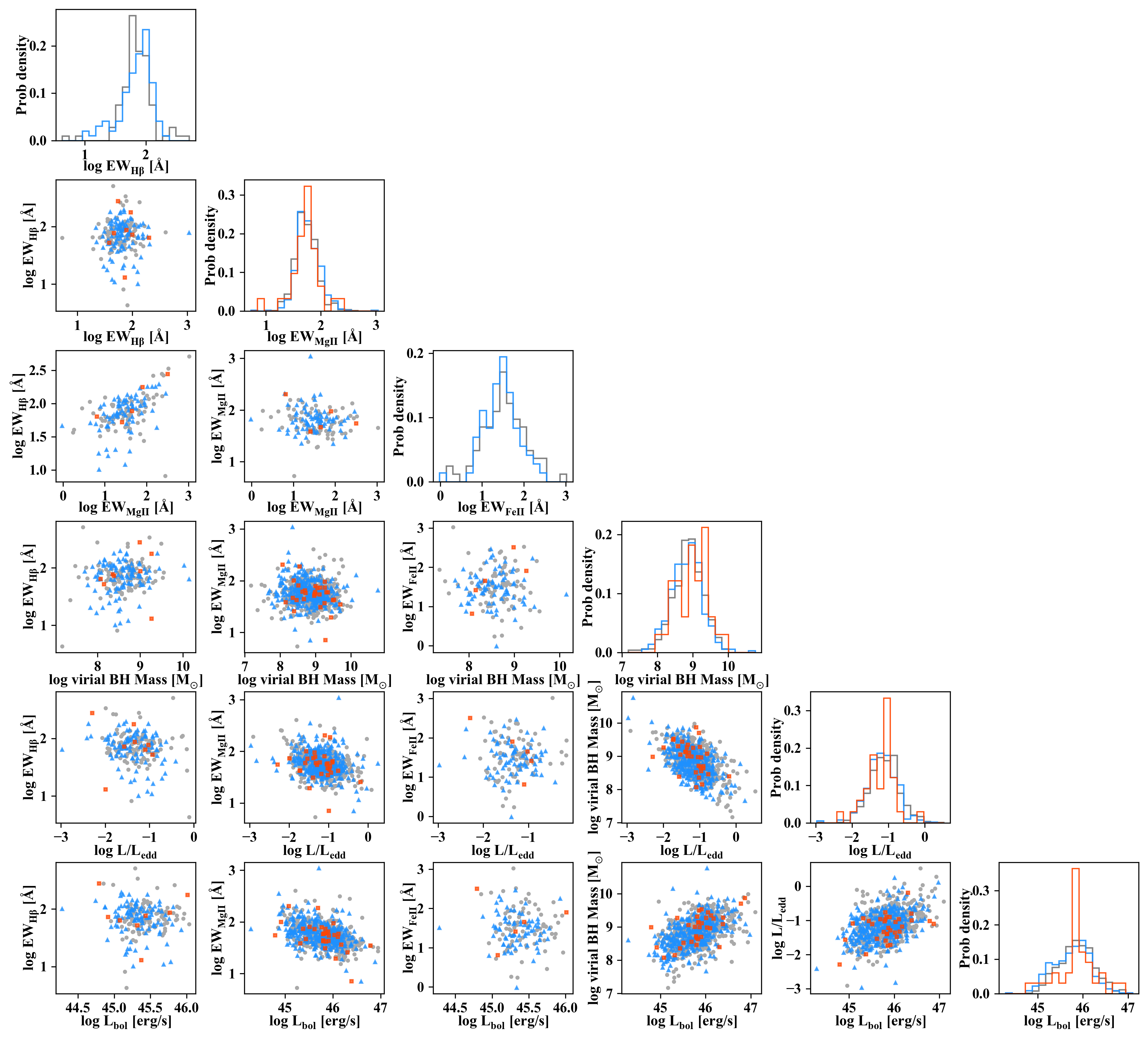}
        \caption{
                \label{fig:property-1} 
         Comparison among the three light curve categories in terms of different physical quasar properties. We require at least 15 objects to plot the histogram. The color scheme and symbols in this plot are the same as in Fig.~\ref{fig:z-vari}. There is no significant difference among the three categories in the properties we examined.       
        }
\end{figure*}

\section{Discussion}\label{sec:disc}

\subsection{Physical dependences of long-term trends}\label{sec:disc_long}

The three categories we introduced in \S\ref{sec:long_var} represent different smoothness in the long-term evolution of the optical light curves for EVQs. We have already demonstrated that this diversity in the long-term trend does not depend on redshift or the short-term variability (Fig.~\ref{fig:z-vari}). Fig.~\ref{fig:property-1} further compares the three categories in terms of the physical properties of quasars measured from optical spectroscopy \citep[][]{Shen_etal_2011}. In all the quantities we examined, there is no significant difference among the three categories. This result suggests that the long-term trends in the optical light curves of EVQs are stochastic, and there is no dominant timescale of physical processes within the observed-frame 16-yr baseline that we can infer to be responsible for the extreme multi-year optical variability. However, the small fraction of objects with monotonic increasing/decreasing light curves over the 16-yr period suggests that such processes must mostly operate on timescales of a few years. For example, if such processes are secular on $\gg 10$ yr timescales, we would expect a much higher fraction of monotonic increasing/decreasing light curves among all EVQs.

\subsection{Microlensing}\label{sec:disc_micro}

\begin{figure} 
        \centering \includegraphics[width=\columnwidth]{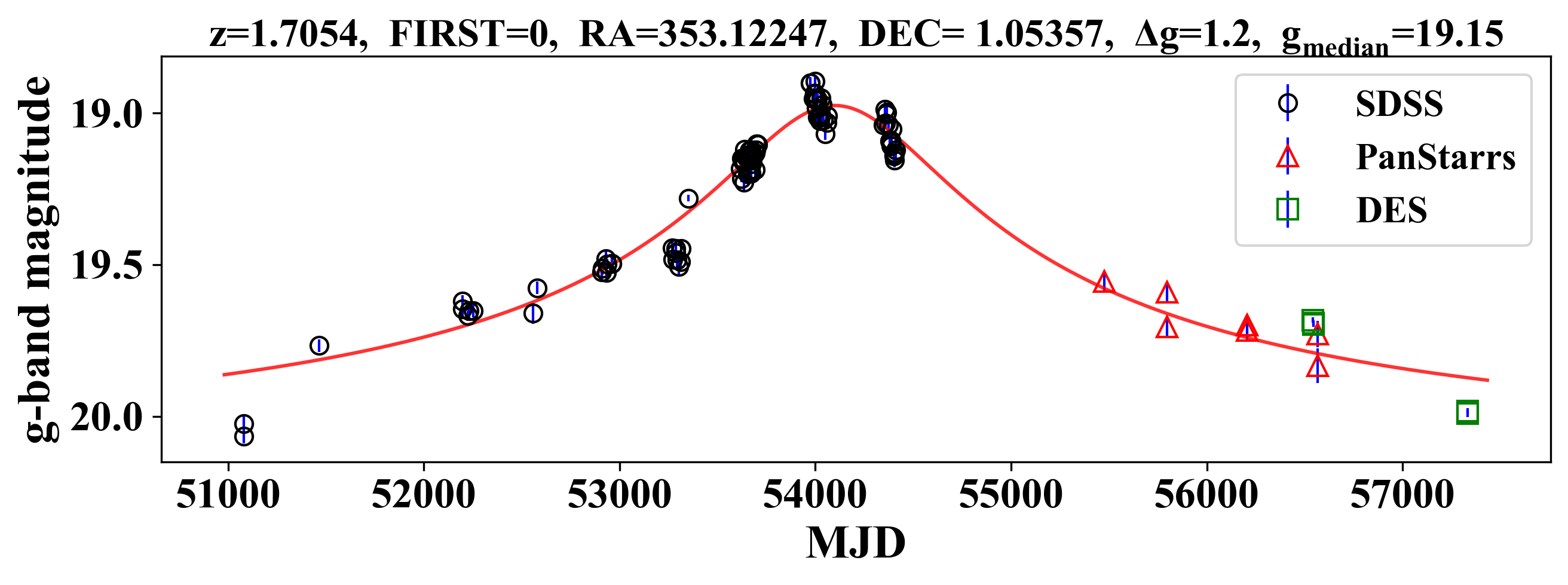}
        \caption{
                \label{fig:mcl} 
                An example of microlensing candidates \citep[ID=104714 in the catalog of][]{Shen_etal_2011}. The red curve represents the best-fit model, and the best-fit parameters are listed in Table \ref{table: mcl info}. The relatively low-amplitude fluctuations around the red curve are likely due to intrinsic quasar variability. 
        }
\end{figure}

\begin{figure*} 
        \centering \includegraphics[width=2\columnwidth]{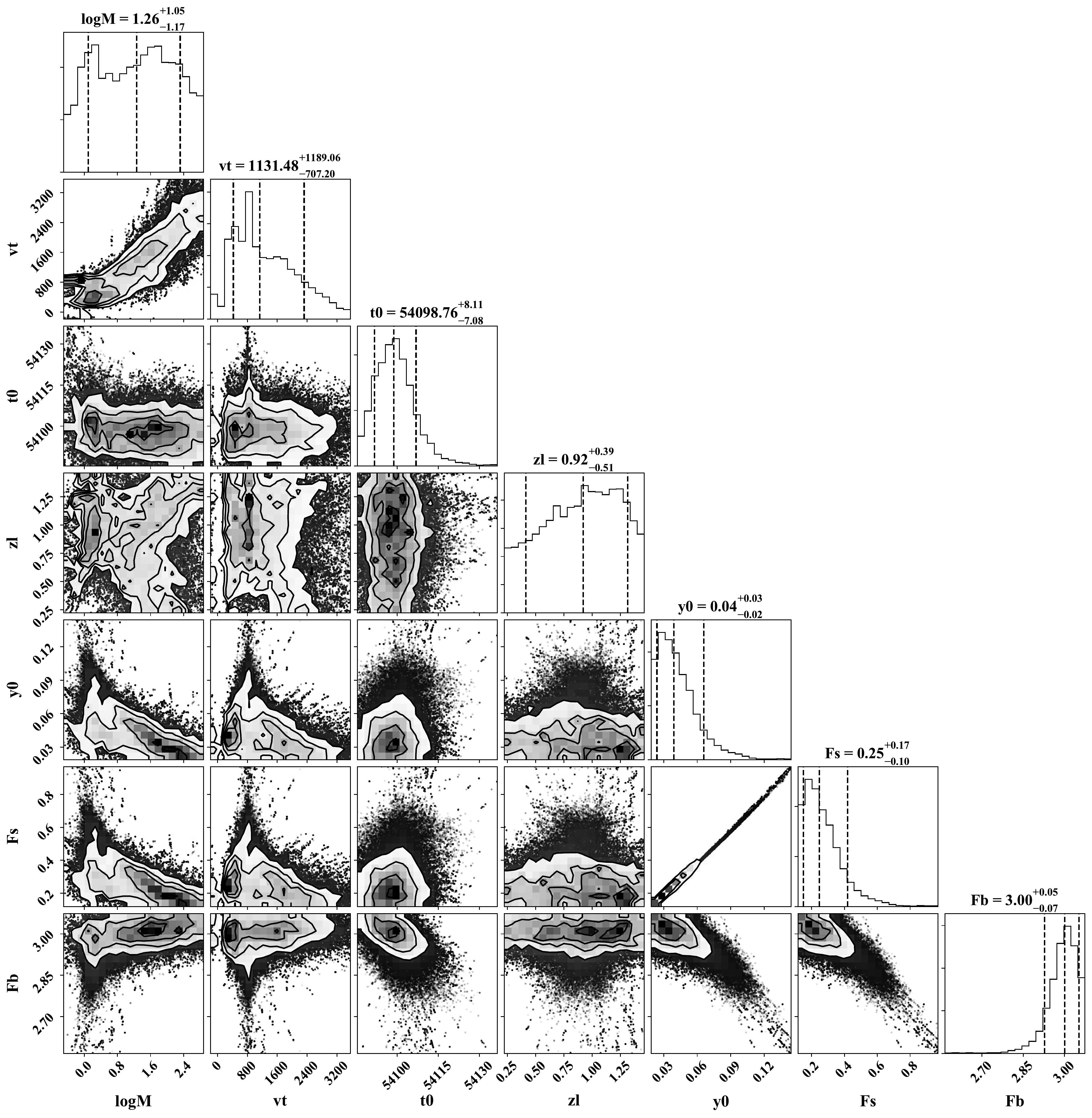}
        \caption{
                \label{fig:corner} 
           Posterior distributions of the parameters from the MCMC fitting of a simple microlensing model to the example light curve (ID=104714) shown in Fig.~\ref{fig:mcl}. The errors shown on the plot are calculated from the 16 and 84 percentiles of the parameter distribution. The 7 fitting parameters are summarized in Table \ref{table: mcl parameter}. 
        }
\end{figure*}

\begin{figure*} 
        \centering \includegraphics[width=2\columnwidth]{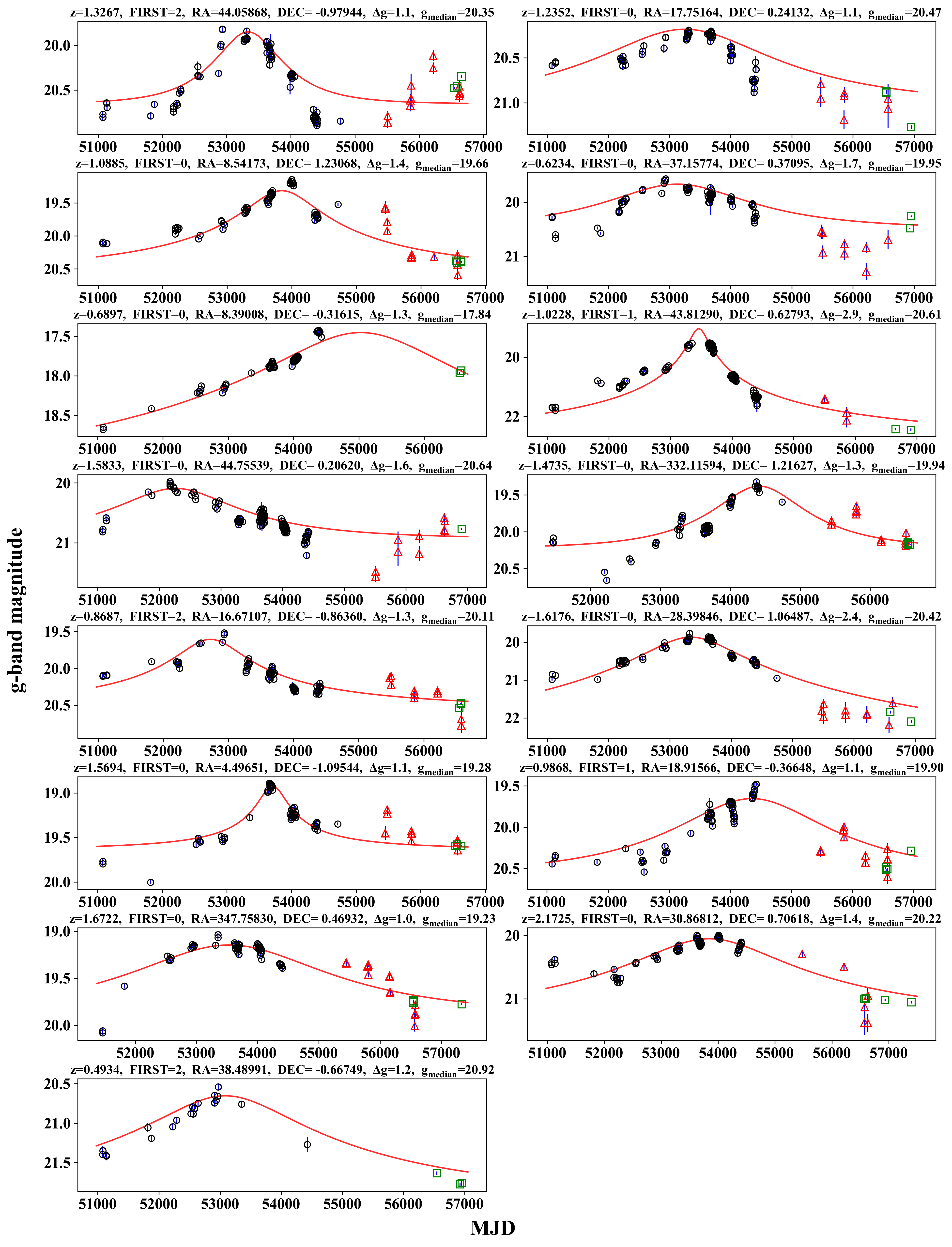}
        \caption{
                \label{fig:mcl_col} 
           Light curves of microlensing candidates (in addition to the one shown in Fig. \ref{fig:mcl}) overplotted with the best-fit models. The color scheme and symbols in this plot are the same as in Fig. \ref{fig:mcl} and the best-fit parameters are listed in Table \ref{table: mcl info}. 
        }
\end{figure*}

\begin{table}
\begin{center}
\caption{Fitting parameters in the point-lens-point-source microlensing model.}
\label{table: mcl parameter}
\begin{tabular}{ c c } \hline
Parameter & Descriptions \\ \hline
M$_l$ & Lens mass\\
v$_t$ & Transverse velocity\\
t$_0$ &  Mid-point epoch\\
z$_l$ & Lens redshift\\
y$_0$ & Impact parameter\\
F$_s$ & Source flux (without lensing)\\
F$_b$ & Background flux (unlensed)\\
\hline
\end{tabular}
\end{center}
\end{table}


Among the $\sim 900$ EVQs, we have identified a handful of objects whose light curve is approximately symmetric and resembles the light curve in a microlensing event \citep[e.g.,][]{Lawrence_etal_2016,Bruce2017,Graham_etal_2017}. Here we fit a simple lens model, the point-source-point-lens model, to 16 candidate microlensing events selected from our EVQ sample based on their bell-shaped light curves in visual inspection. In this model we assume that both the source and the lens are point-like, and the fits are performed with the MCMC package implemented in \citet{MCMC}.

Following \citet{Bruce2017}, the flux magnification associated with the microlensing event is given by
\begin{equation}
    \mu = \frac{y(t)^2+2}{y(t)\sqrt{y(t)^2+4}}\ ,
\end{equation} 
where $y(t)$ is the distance between the source and the lens in units of the Einstein radius of the lens $\theta_{E}$, which is given by
\begin{equation}
 \theta_E = \left(\frac{4GM_l}{c^2}\frac{D_{ls}}{D_lD_s}\right)^{1/2}\ ,
\end{equation}
where $D_l$, $D_s$ and $D_{ls}$ are the angular diameter distances for the lens, source and between the lens and source, respectively. These distances are calculated with the source redshift and lens redshift, and y(t) can be calculated using 
\begin{equation}
        y(t) = \sqrt{y_0^2+\left(\frac{t-t_0}{t_E}\right)^2},\ \ t_E= \frac{D_l\theta_E}{v_{\perp}}\ .
\end{equation}
Here $y_0$ is the impact parameter at $t_0$, and $v_{\perp}$ is the transverse velocity of the lens relative to the observer-source line of sight. The final observed flux is given by $ F(t) = \mu(t)F_s + F_b$, where we also take into consideration the unlensed background flux. The seven free parameters in the model are summarized in Table \ref{table: mcl parameter}. 

Fig.~\ref{fig:mcl} shows an example of the microlensing candidates with the best-fit model overplotted. The posterior distributions of the fitting parameters are shown in Fig.~\ref{fig:corner}. We summarize the fitting results for all 16 microlensing candidates in Table \ref{table: mcl info}. Because of the stochastic intrinsic variability of quasars (not accounted for in the fit) and small measurement uncertainties of the light curve, the model fit is generally not a good fit in terms of the reduced $\chi^2$ value. But the overall agreement between the model and the observed light curve suggests that microlensing is a viable explanation for the observed extreme optical variability. The rest of the microlensing candidates are plotted with their best-fit models in Fig. \ref{fig:mcl_col}.

For the simulated EVQ light curves described in \S\ref{sec:long_var_drw} (either with $SF_{\infty} = 0.2$ mag and $\tau=200$ days, or with $SF_{\infty} = 0.4$ mag and $\tau=600$ days), we identify light curves with relatively smooth and symmetric profiles and fit with a microlensing model. The resulting reduced $\chi^2$ values are all very large ($\gg$100 if assuming magnitude errors comparable to those of the real data) and do not represent those of the microlensing events from real data. It is thus unlikely that many of the 16 microlensing events are due to stochastic quasar variability.

\begin{table*}
\begin{center}
\caption{Best-fit parameters of the point-source-point-lens model for 16 microlensing candidates. The first column is the object index (starting from zero) from the SDSS DR7 quasar catalog in \citet{Shen_etal_2011}. z$_s$ is the source redshift. Because of the stochastic intrinsic variability of quasars (not accounted for in the fitting) and small measurement uncertainties of the light curve data, the model fit is generally not a good fit in terms of the reduced $\chi^2$ value. The errors are calculated from the 16 and 84 percentiles of the parameter distribution. }
\label{table: mcl info}
\begin{tabular}{ c c c c c c c c c c c} \hline
\\[-1em]
ID & z$_s$ & logM$_l$ [M$_{\sun}$] & v$_t$ [km/s] & t$_0$ [MJD] &  z$_l$ & y$_0$ [$\theta_E$] & F$_s$ [10$^{-5}$ Jy] & F$_b$ [10$^{-5}$ Jy] & $\chi_{\rm red}^2$\\ 
\\[-1em]
\hline
\\[-1em]
\\[-1em]
684 & 1.5694 &  0.00$^{+1.36}_{-1.19}$ & 1178.62$^{+2050.29}_{-781.33}$ & 53707.78$^{+6.23}_{-7.72}$ & 0.74$^{+0.53}_{-0.44}$ & 0.07$^{+0.05}_{-0.03}$ & 0.33$^{+0.26}_{-0.15}$ & 4.75$^{+0.09}_{-0.21}$ & 19.03\\
\\[-1em]
\\[-1em]
\\[-1em]
1296 & 0.6897 & 0.97$^{+0.85}_{-0.60}$ & 874.68$^{+783.49}_{-444.20}$ & 55029.42$^{+6.53}_{-6.88}$ & 0.32$^{+0.16}_{-0.14}$ & 0.10$^{+0.04}_{-0.04}$ & 3.50$^{+1.57}_{-1.27}$ & 2.38$^{+0.59}_{-0.78}$ & 23.95\\
\\[-1em]
\\[-1em]
\\[-1em]
1322 & 1.0885 & 1.16$^{+1.09}_{-1.04}$ & 1047.71$^{+1317.13}_{-615.83}$ & 53846.21$^{+7.30}_{-7.49}$ & 0.58$^{+0.30}_{-0.35}$ & 0.04$^{+0.03}_{-0.02}$ & 0.20$^{+0.16}_{-0.08}$ & 1.80$^{+0.04}_{-0.07}$ & 27.42\\
\\[-1em]
\\[-1em]
\\[-1em]
2657 & 0.8687 & 1.06$^{+1.21}_{-1.15}$ & 1129.54$^{+1722.38}_{-703.87}$ & 52724.45$^{+10.51}_{-10.75}$ & 0.41$^{+0.26}_{-0.25}$ & 0.05$^{+0.04}_{-0.02}$ & 0.15$^{+0.14}_{-0.07}$ & 1.98$^{+0.04}_{-0.08}$ & 12.86\\
\\[-1em]
\\[-1em]
\\[-1em]
2820 & 1.2352 & -0.26$^{+0.71}_{-0.90}$ & 873.08$^{+1323.87}_{-492.37}$ & 53250.49$^{+15.93}_{-14.60}$ & 0.64$^{+0.37}_{-0.40}$ & 0.40$^{+0.09}_{-0.22}$ & 1.00$^{+0.33}_{-0.62}$ & 0.42$^{+0.53}_{-0.31}$ & 32.31\\
\\[-1em]
\\[-1em]
\\[-1em]
2988 & 0.9868 & -0.46$^{+0.59}_{-0.68}$ & 806.46$^{+914.11}_{-355.86}$ & 54353.80$^{+14.70}_{-15.29}$ & 0.61$^{+0.26}_{-0.37}$ & 0.45$^{+0.01}_{-0.04}$ & 2.08$^{+0.07}_{-0.23}$ & 0.08$^{+0.23}_{-0.06}$ & 38.90\\
\\[-1em]
\\[-1em]
\\[-1em]
4780 & 1.6176 & 1.90$^{+0.75}_{-0.89}$ & 1137.66$^{+972.29}_{-663.98}$ & 53341.72$^{+5.36}_{-5.48}$ & 0.85$^{+0.36}_{-0.45}$ & 0.02$^{+0.01}_{-0.01}$ & 0.09$^{+0.05}_{-0.03}$ & 0.00$^{+0.00}_{-0.00}$ & 13.59\\
\\[-1em]
\\[-1em]
\\[-1em]
5212 & 2.1725 & 0.12$^{+0.79}_{-0.80}$ & 931.13$^{+1006.54}_{-547.27}$ & 53841.80$^{+13.93}_{-13.26}$ & 1.03$^{+0.66}_{-0.59}$ & 0.23$^{+0.08}_{-0.09}$ & 0.69$^{+0.30}_{-0.27}$ & 0.42$^{+0.17}_{-0.23}$ & 25.58\\
\\[-1em]
\\[-1em]
\\[-1em]
6301 & 0.6234 & -0.48$^{+0.60}_{-0.64}$ & 834.84$^{+1115.79}_{-299.72}$ & 53128.67$^{+7.79}_{-8.70}$ & 0.38$^{+0.17}_{-0.24}$ & 0.50$^{+0.01}_{-0.05}$ & 2.25$^{+0.05}_{-0.17}$ & 0.06$^{+0.26}_{-0.05}$ & 52.02\\
\\[-1em]
\\[-1em]
\\[-1em]
6525 & 0.4934 & 0.86$^{+1.28}_{-0.95}$ & 1236.98$^{+2962.67}_{-860.94}$ & 53084.98$^{+54.80}_{-49.86}$ & 0.23$^{+0.17}_{-0.16}$ & 0.16$^{+0.09}_{-0.07}$ & 0.26$^{+0.18}_{-0.12}$ & 0.33$^{+0.09}_{-0.14}$ & 13.25\\
\\[-1em]
\\[-1em]
\\[-1em]
7332 & 1.0228 & 1.22$^{+0.82}_{-1.04}$ & 1369.02$^{+1041.73}_{-682.78}$ & 53459.76$^{+1.09}_{-1.07}$ & 0.60$^{+0.19}_{-0.26}$ & 0.01$^{+0.01}_{-0.01}$ & 0.12$^{+0.07}_{-0.04}$ & 0.14$^{+0.02}_{-0.02}$ & 49.09\\
\\[-1em]
\\[-1em]
\\[-1em]
7384 & 1.3267 & -1.17$^{+0.91}_{-0.83}$ & 1108.33$^{+1686.66}_{-633.57}$ & 53334.22$^{+9.92}_{-9.62}$ & 0.61$^{+0.44}_{-0.40}$ & 0.52$^{+0.01}_{-0.01}$ & 1.97$^{+0.02}_{-0.03}$ & 0.02$^{+0.03}_{-0.01}$ & 35.23\\
\\[-1em]
\\[-1em]
\\[-1em]
7494 & 1.5833 & -0.31$^{+0.81}_{-0.85}$ & 1156.78$^{+1744.91}_{-672.99}$ & 52261.23$^{+13.01}_{-12.73}$ & 0.83$^{+0.48}_{-0.50}$ & 0.37$^{+0.08}_{-0.08}$ & 0.99$^{+0.29}_{-0.25}$ & 0.55$^{+0.24}_{-0.27}$ & 17.11\\
\\[-1em]
\\[-1em]
\\[-1em]
102024 & 1.4735 & -0.79$^{+0.74}_{-0.82}$ & 890.46$^{+1287.15}_{-454.18}$ & 54412.89$^{+22.41}_{-20.86}$ & 1.06$^{+0.26}_{-0.53}$ & 0.45$^{+0.04}_{-0.09}$ & 2.48$^{+0.34}_{-0.63}$ & 0.48$^{+0.61}_{-0.34}$ & 48.33\\
\\[-1em]
\\[-1em]
\\[-1em]
103976 & 1.6722 & -0.22$^{+0.76}_{-0.72}$ & 1103.88$^{+1535.60}_{-590.66}$ & 53533.76$^{+15.42}_{-15.03}$ & 0.84$^{+0.48}_{-0.48}$ & 0.55$^{+0.02}_{-0.05}$ & 3.83$^{+0.18}_{-0.47}$ & 0.22$^{+0.44}_{-0.17}$ & 26.57\\
\\[-1em]
\\[-1em]
\\[-1em]
104714 & 1.7054 & 1.26$^{+1.05}_{-1.17}$ & 1131.48$^{+1189.06}_{-707.20}$ & 54098.76$^{+8.11}_{-7.08}$ & 0.92$^{+0.39}_{-0.51}$ & 0.04$^{+0.03}_{-0.02}$ & 0.25$^{+0.17}_{-0.10}$ & 3.00$^{+0.05}_{-0.07}$ & 13.87\\
\\[-1em]
\hline
\end{tabular}
\end{center}
\end{table*}

\section{Conclusions}\label{sec:con}

In this work we studied the 16-yr optical light curves for a large sample of EVQs using public data from multiple imaging surveys. These light curves allowed us to explore the diversity in the long-term trends of these EVQs and their implications for accretion disk physics. Our main findings are the following:

\begin{enumerate}
    \item We classify the light curves into three categories with different levels of smoothness in their long-term (multi-year) trends over the 16-yr baseline: monotonic decreasing or increasing (3.7\%), a single broad peak and/or dip (56.8\%) and more complex patterns (39.5\%). The observed fractions of relatively smoother long-term light curves (monotonic and single peak/dip) are higher than those predicted by stochastic quasar variability based on the damped random walk model. 
    \item Despite the difference in their long-term light curve trend, EVQs in the three categories have indistinguishable distributions in their quasar properties, such as short-term (seasonal) variability, redshift, BH mass and Eddington ratio.
    \item We measure the rms-mean flux relation using the light curves of these EVQs. The short-term rms variability (in magnitude) decreases as the seasonal-average flux increases, which is different from X-ray variability studies where the rms flux (in linear units) scales with the mean flux. This result suggests different origins for the short-term variability and the long-term evolution in the optical light curve. 
    \item We presented a sample of 16 EVQs where the relatively symmetric light curve can be reasonably well produced by a simple microlensing model, offering an alternative explanation for the extreme variability over multi-year timescales.  
\end{enumerate}

Our results provide further insight on the nature of extreme optical variability in quasars. The long-term evolution of the optical light curve is stochastic, and does not depend on the physical properties of EVQs. No matter what process is driving the extreme multi-year optical variability from the accretion disk, it mostly operates on timescales of a few years as constrained by the observed-frame 16-yr baseline. For example, if such processes are gradual over $\gg 10$ yrs timescales, we should expect a much higher fraction of monotonic trends among the light curves than what is observed. The diversity and general complexity in the shapes of these decade-long light curves also rule out a single transient event (such as microlensing and tidal tidal disruption events) as the dominant mechanism for the observed extreme optical variability \citep[e.g.,][]{Dexter_etal_2019}.  

The lack of a linear relation between the short-term optical rms flux and the seasonal-average flux, as observed for X-ray variability, suggests different mechanisms that drive the short-term flickering and long-term extreme variability from the accretion disk in EVQs. We suggest that the extreme optical variability is more likely driven by in situ instabilities in the accretion disk modulated by their propagation/evolution timescales \citep[e.g.,][]{Graham_etal_2019}, whereas the moderate short-term  flickering can be generally driven by reprocessing of the variable X-ray emission from the corona \citep[e.g.,][]{Frank_2002}. 

\section*{Acknowledgements}

We thank the referee for useful comments that greatly improved the manuscript, and Yan-Fei Jiang, Jennifer I-Hsiu Li for useful discussion. YS and QY acknowledge support from an Alfred P. Sloan Research Fellowship (YS) and NSF grant AST-1715579.

\bibliographystyle{mnras}
\bibliography{refs,ref} 

\bsp	
\label{lastpage}

\end{document}